\documentclass[final,3p,times,twocolumn,authoryear]{elsarticle}

\usepackage{amsmath}
\usepackage{amssymb}
\usepackage{xcolor}

\journal{New Astronomy}

\begin{document}

\begin{frontmatter}



\title{Measurable Parameter Combinations of Environmentally-dephased EMRI Gravitational-Wave Signals}


\author[inst1]{Marco Immanuel B. Rivera}

\affiliation[inst1]{organization={Department of Physics, University of Rome Tor Vergata},%Department and Organization
            addressline={Via Della Ricerca Scientifica 1}, 
            city={Rome},
            postcode={00133},
            country={Italy}}

\author[inst2]{Reinabelle C. Reyes}
% \author[inst1,inst2]{Author Three}

\affiliation[inst2]{organization={National Institute of Physics, University of the Philippines},%Department and Organization
            addressline={Diliman}, 
            city={Quezon City},
            postcode={1101}, 
            state={NCR},
            country={Philippines}}

\begin{abstract}
% 1. We investigate XX by doing YY.
% 2. Motivation
% 3. Method explanation
% 4. Results
% 5. Importance

The future space-borne Laser Interferometer Space Antenna (LISA) is expected to detect gravitational waves (GW) from Extreme Mass Ratio Inspiral (EMRI) binaries which may live in nontrivial environments such as accretion disks. 
In this work, we apply the Fisher matrix Principal Component Analysis (PCA) method to assess how well LISA observations can jointly constrain the source parameters and environmental densities around EMRIs. Specifically, we calculate the Fisher matrix from the post-Newtonian parameters of an EMRI binary embedded in a fluid with a constant density profile. We determine that the most dominant measurable parameter combination is dominated by contributions from environmental effects, namely, gravitational drag, accretion, and gravitational pull (in order of contribution). The proposed reparameterization of the PN parameters can be used to improve the power and efficiency of future detection and parameter estimation methods.
\end{abstract}


\begin{highlights}

% Rewrite highlights - to also emphasize the contribution and new results
% First application of PCA to environmental…  
% Result: first PC - environmental effects
% Applications

\item We apply the Fisher matrix Principal Components Analysis (PCA) for gravitational waves imprinted with non-trivial environmental effects. For an Extreme-Mass Ratio Inspiral (EMRI) binary system, we determine the most dominant measurable parameter combination, which is a combination of the effects of gravitational drag, accretion, and gravitational pull (in order of contribution). 

\item We propose a reparameterization of the post-Newtonian (PN) parameters encapsulating the environmental effects to improve the efficiency of future detection and parameter estimation methods.

\item Our analysis indicates that determining measurable parameter combinations using methods such as Fisher Matrix PCA will be relevant for strategies in constraining the environments around EMRIs in general, and this approach can be readily expanded to probe other systems in the LISA range.

% \item Using Principal Components Analysis (PCA), we construct parameter combinations alternative to the chirp mass, especially for gravitational waveforms with dephasing from environmental effects. The derived principal components allow for faster parameter estimation by tracking the new parameter combinations instead of the chirp mass or individual masses.

% \item The first principal component effectively replaces the chirp mass as the most accurate parameter combination for these environmentally-dephased binaries. Compared with chirp mass, it can be more precise in distinguishing between less massive and more asymmetric binaries, but it cannot distinguish between binaries in varying environmental densities.
\end{highlights}


\end{frontmatter}


\section{Introduction}
\label{sec:intro}
Extreme mass-ratio inspiral (EMRI) binaries consist of a stellar-mass black hole or other compact object orbiting a supermassive black hole in the centers of galaxies. They are among the most interesting sources for future space-borne gravitational-wave detectors such as Laser Interferometer Space Antenna (LISA). These binaries may live in non-trivial astrophysical environments in the form of gas or dark matter, which affect the orbital evolution of the binary and therefore the emission of gravitational waves (GW). 

% Minor: LIGO-Virgo data has now been used to provide updated constraints on environments, https://arxiv.org/abs/2309.05061. perhaps the authors want to update the introduction with this reference.

In particular, dark matter densities in cold dark matter spikes can range from $10^{-15}$ to $10^{-9}\mbox{ kg m}^{-3}$, while the baryonic densities in accretion disks can reach up to $10^2-10^3 \mbox{ kg m}^{-3}$~\citep{barausse2014,Santoro2023}. The most dominant environmental imprints on the GW signals come from three mechanisms: accretion onto the primary black hole, gravitational pull (i.e., the matter in accretion disks or in other structures exerts a gravitational pull on the compact objects), and gravitational drag (i.e., the gravitational interaction of the compact objects with their own wake in the matter medium). These effects can contribute to a small, but potentially observable, change in the gravitational-wave phase~\citep{Cardoso2020,barausse2014}. Recently, \cite{Santoro2023} analyzed events from the first GW catalog from LIGO-Virgo (GWTC-1) to look for potential imprints of environmental effects; they found no significant evidence of their presence, confirming that these sources can be treated as in-vacuum isolated binary systems, while also demonstrating the feasibility of determining constraints on environmental densities from real-world data.

% analysis of Ligo-Virgo data [cite paper]; contrast, vacuum environments vs. emri na mas may environments.

% Add paragraph discussing ”traditional” Fisher Matrix Analysis [cite relevant papers + Cardoso et al 2021] -> explain/contrast with Fisher + PCA: how this is modified/expanded to get parameter combinations.

The Fisher matrix is a popular and convenient method to assess the ability of future GW detectors to constrain non-trivial astrophysical environments, such as in EMRIs. The Fisher matrix is defined by an inner (noise-weighted) product of derivatives of the waveform with respect to its parameters. By calculating the inverse of the Fisher matrix, one gets an estimate of the covariance matrix and consequently, the uncertainties of parameters of interest~\citep{finn1993observing,Cutler1994,Vallisneri2008,Cardoso2020,Cardoso2021}. Specifically, Fisher matrix analysis has been applied to show that LISA observations can constrain the environmental density around EMRI binaries down to densities typical of dark matter, with the tightest constraints coming from the effect of gravitational drag~\citep{Cardoso2020,SPP-2021-1C-02}. 
% This indicates that environmental effects could impact the interpretation of gravitational-wave signals observed by LISA. 

The Fisher matrix analysis, while convenient, has recognized limitations. It has been pointed out that the inverse of the Fisher matrix is only a good estimate of the covariance matrix in the limit of high signal-to-noise ratio (SNR)~\citep{Vallisneri2008}. There are also cases where the Fisher matrix is ill-conditioned and numerical inversion routines introduce large numerical errors~\citep{Berti2005,Pai2013,Ohme2013}. An alternative and supplementary approach, which sidesteps these issues, performs Principal Component Analysis (PCA) on the Fisher matrix instead of taking its inverse. Notably, rather than being limited to the original physical parameters, this method also determines the combinations of physical parameters that are most effectively measured by the detector, which are-- precisely-- the calculated principal components. Moreover, the uncertainties of these measurable parameter combinations can be derived from the eigenvalues calculated from the PCA.

The Fisher matrix PCA approach has previously been successfully applied to determine measurable parameter combinations for compact binary systems and study the systematic effect of various spin contributions to the inspiral waveform model, assuming environmental effects are not present~\citep{Ohme2013,Pai2013,Pannarale2014,Lee2022}. In this vacuum case, it has been confirmed that the dominant parameter of the inspiral waveform is the chirp mass, and the next dominant parameter was found to be a combination of the spin and the symmetric mass ratio~\citep{Ohme2013}. In this work, we extend the application of this framework to EMRIs in non-trivial astrophysical environments to assess how well LISA observations can constrain the source parameters and environmental densities around EMRIs. 

A range of studies have shown that environmental effects are expected to be observable and have to be taken into account in the interpretation of GW signals from EMRIs, especially for systems with confirmed EM counterparts. In particular, studies have investigated how they (1) can change the gravitational waveform \citep{Barausse2007,Kocsis2011,yunes2009ppE}, (2) can affect precision parameter estimation \citep{barausse2014,Barausse2015,Chen2020}, (3) can affect EMRI detection rates (either in EM or GW) \citep{Caputo2020,McGee2020}, or (4) can be inferred and estimated from GW observations \citep{Cardoso2020,speri2023probing,cole2023distinguishing,SPP-2021-1C-02}. We contribute to this body of work by presenting measurable parameter combinations that include environmental parameters and studying degeneracies in the parameter space. Our results can serve as basis for reparameterizations for improving the efficiency of stochastic sampling techniques used for parameter estimation, such as the Markov Chain Monte Carlo (MCMC) method, as demonstrated by~\cite{Lee2022}. More generally, they can inform strategies for more efficient detection and parameter estimation, such as providing basis for optimizing the choice of template spacing in matched filtering, which is key to improving computational efficiency~\citep{Owen1996,brown2012detecting,Ohme2013}. 

% Explicitly state why parameter combinations are useful. Even if they do not necessarily correspond to “physical observables” (like chirp mass, spin), relevant “experimentally” because they are the most constrained. 

We note that while this initial investigation is limited by many simplifying assumptions (e.g., we consider only EMRI binaries with quasi-circular orbits on a fixed plane with no effects from eccentricity and precession, adopt an analytic PN waveform approximant calibrated to lower mass binaries, assume that the PN parameters by themselves are free and independent of one another, and obtain the results using a uniform constant environmental density), it can still serve as a good baseline for future studies on studying the waveform parameter space.

The paper is divided as follows. In Section \ref{sec:method} we outline the waveforms used to describe the inspiralling binary, including the corrections from environmental effects. We also outline here the Fisher matrix and Principal Components Analysis (PCA) from which we obtain the parameter combinations. In Section \ref{sec:measurable} we describe the target system for which we apply the PCA and construct the measurable parameter combinations for both vacuum and environmentally-dephased waveforms. We summarize and conclude in Section \ref{sec:conclu}.

\section{Waveform and Methodology}
\label{sec:method}

For this section and the rest of the paper, we use geometric units where $G = c = 1$, unless otherwise specified.

\subsection{Vacuum Waveform Model}
\label{subsec:method_vacuum}

For our vacuum waveform model, we use the IMRPhenomB waveform approximant~\cite{Ajith2011}, which is part of the class of inspiral-merger-ringdown phenomenological waveforms derived by matching analytical PN waveforms to numerical relativity waveforms. For a binary black hole (BBH) source with component masses $m_1$ and $m_2$ and spin angular momentum magnitudes $S_1$ and $S_2$, the frequency-domain waveform is given by
\begin{equation}
    \tilde{h}(f) = \mathcal{A}(f) e^{i\Psi(f)}
    \label{eq:htilde_vac}
\end{equation}
with amplitude $\mathcal{A}$ and phase $\Psi$, defined as follows:
\begin{equation}
    \mathcal{A}(f) = \frac{\sqrt{15}}{2} \frac{1}{\sqrt{30} \pi^{2/3}} \frac{\mathcal{M}}{d_L} f^{-7/6}
    \label{eq:IMRPhenomB_amp}
\end{equation} 
where $\mathcal{M}=M\eta^{3/5}$ is the chirp mass, $M=m_1+m_2$ as the total mass and $\eta = (m_1 m_2)/(m_1+m_2)^2$ as the symmetric mass ratio ($\eta=0.25$ for an equal-mass binary), and $d_L$ is the luminosity distance to the source. We note that this Newtonian amplitude is valid in the post-Newtonian regime, which is the focus of our analysis.

The phase $\Psi$ is given by the sum of PN phase terms
\begin{equation}
    \Psi(f) \equiv \sum_{k=0}^8 \Psi_k = \frac{3}{128 \eta} (\pi M f)^{-5/3} \sum_{k=0}^8 \phi_k \nu^k
    \label{eq:IMRPhenomB_phase}
\end{equation}
where $\nu = (\pi M f)^{1/3}$ is also known as the reduced GW frequency and the coefficients $\phi_k$'s are:
\begin{eqnarray}
 \phi_0 &=& 1 \nonumber\\
 \phi_1 &=& 0 \nonumber \\ 
 \phi_k &=& \psi_k^0(\chi) + \sum_{m=1}^3 \sum_{n=0}^N x_k^{(mn)} \eta^m \chi^n \textrm{ for } k \in \{2,...,7\} \nonumber \\ 
 \phi_8 &=& 2\pi t_c f. \nonumber 
\end{eqnarray}
Here, the values of $\psi_k^0(\chi)$ and $x_k^{(mn)}$ are outlined in Table I of \cite{Ajith2011}, $\chi = (m_1\chi_1 + m_2\chi_2)/M$ is the effective spin parameter, where $\chi_{1,2} = cS_{1,2}/G m_{1,2}^2$ is the dimensionless spin parameter of each source component, and $t_c$ is the time at coalescence.

%We note that these waveforms were calibrated to low-mass binaries but are adopted here for their simplicity \textit{[NOTE implication for using with our EMRI target with $m_1=10^5$ and $m_2=10$]}.

\subsection{Environmental Dephasing}
\label{subsec:method_dephasing}

To model the effect of astrophysical environments, we follow the treatment of \cite{barausse2014} and \cite{Cardoso2020} and approximate the environmental effects on the vacuum waveform by a dephasing factor akin to a Parametrized Post-Einsteinian (PPE) phase term \citep{yunes2009ppE}. For a constant environmental density $\rho_0$, the environmentally-dephased waveform is given by
\begin{equation}
   \tilde{h}_{\rm env}(f) = \mathcal{A}(f) e^{i(\Psi(f) + \Psi_{\rm env}(f))}
   \label{eq:htilde_env}
\end{equation}
with $\Psi_{\rm env} = \Psi_{-11} + \Psi_{-9} + \Psi_{-6}$; here, each term denotes the effect of gravitational drag, accretion, and gravitational pull, respectively, given by
\begin{eqnarray}
    \Psi_{-11} &=& - \frac{3}{128 \eta} (\pi M f)^{-5/3} \frac{(1-3\eta)}{\eta^{3}} M^2\rho_0 \nu^{-11} \\
    \Psi_{-9} &=&  - \frac{3}{128 \eta} (\pi M f)^{-5/3} \frac{1}{\eta} M^2\rho_0 \nu^{-9}  \\
    \Psi_{-6} &=& \frac{3}{128 \eta} (\pi M f)^{-5/3} \pi^2 M^2\rho_0 \nu^{-6}. 
\end{eqnarray}
Effectively, we add these 3 PN parameters with negative exponents to the 9 PN parameter phase expansion of the IMRPhenomB approximant.

\subsection{Fisher Matrix and Principal Component Analysis}
\label{subsec:method_PCA}

We follow the approach of using Fisher Matrix and Principal Component Analysis (PCA) in \cite{Ohme2013} which allows us to determine the most accurately measurable parameter combinations from observed gravitational waveforms. 

We start by calculating the Fisher matrix elements using the derivatives of the waveform with respect to the PN parameters $\theta_i$ given by the overlap integral
\begin{eqnarray}
    \Gamma_{ij} &=& \mathcal{O}\left(\frac{\partial \tilde{h}}{\partial \theta_i} \left| \frac{\partial \tilde{h}}{\partial \theta_j} \right) \right. \nonumber \\
    &=& 4 \Re \int_{f_{\rm min}}^{f_{\rm max}} \frac{df}{S_n(f)} \frac{\partial \tilde{h}}{\partial \theta_i} \left(\frac{\partial \tilde{h}}{\partial \theta_j}\right)^*
    \label{eq:FisherMatrix}
\end{eqnarray}
where $\Re$ denotes taking the real part, $^*$ denotes complex conjugation, and $S_n(f)$ is the noise spectral density of LISA \citep{Robson2019}. For the bounds of the integral, we adopt the LISA frequency bounds from \cite{Berti2005} and \cite{Cardoso2020} given by
\begin{eqnarray}
    f_{\rm min} &=& \max \left[10^{-5}, f_{\rm ref}\right]~{\rm Hz} \\
    f_{\rm max} &=& \min [1, ~f_{1}] ~{\rm Hz}
\end{eqnarray}
where

\begin{equation}
    f_{\rm ref} = 4.15 \times 10^{-5}~{\rm Hz} \left(\frac{\mathcal{M}}{10^6 M_{\odot}}\right)^{-5/8} \left(\frac{T_{{\rm obs}}}{\rm yr}\right)^{-3/8}.
\end{equation}

\noindent Here, $f_{1}$ is the transition frequency from inspiral to merger in the IMRPhenomB formulation \citep{Ajith2011}, and $T_{\text{obs}}$ is the length of observation time for the LISA mission, which we set to 4 years. 

%The bounds $10^{-5}$ [Hz] and $1$ [Hz] depend on the design sensitivity of LISA, which changes frequently to keep up with upgrades \cite{Berti2005, Robson2019}. 
%%RR Note: Details taken from SPP paper

As in \cite{Ohme2013}, we calculate the derivatives in Equation \eqref{eq:FisherMatrix} with respect to the PN parameters themselves, treating them as independent and orthogonal bases in their own parameter space. For the vacuum waveform, the Fisher matrix is an 8 $\times$ 8 matrix because one of the PN parameters is zero ($\Psi_1 = \phi_1 = 0$), i.e., $\theta_i =$ \{$\Psi_0$, $\Psi_2$, $\Psi_3$,  \ldots, $\Psi_8\}$. For the environmentally-dephased waveform, this is an 11 $\times$ 11 matrix, with $\theta_i = \{\Psi_{-11}$, $\Psi_{-9}$, $\Psi_{-6}$, $\Psi_0$, $\Psi_2$, $\Psi_3$, \ldots, $\Psi_8\}$. In both cases, we marginalize over the nuisance parameters, phase and time at coalescence (which appear in $\Psi_5$ and $\Psi_8$, respectively), reducing the Fisher matrix to $6 \times 6$ in the vacuum case and $9 \times 9$ in the environmentally-dephased case \citep{long2015fisher}.

Next, we get the eigenvalues and eigenvectors of this marginalized Fisher matrix, which we will denote by $\{\lambda_i\}$ and $\{\psi_i\}$, respectively. We also get the theoretical spread in the 90\% confidence interval $\Delta \mu_i = \left( \chi_k^2 / \lambda_i \right)^{1/2}$, assuming all PN coefficients to be free and independent parameters \citep{Ohme2013,Baird2013}. Here, $\chi_k^2$ is the $\chi^2$ value for which the probability of obtaining that value or less in a $\chi^2$ distribution with $k$ degrees of freedom is 90\%. For the system we will consider, the number of degrees of freedom $k=3$ ($\chi_3^2 = 6.25$) in the vacuum case (two component masses and the effective spin parameter) and $k=4$ ($\chi_4^2 = 7.78$) for the environmentally-dephased case (with the addition of the environmental density). This corresponds to the distance criterion
\begin{eqnarray}
||h(\theta) - h(\hat{\theta})||^2 &\equiv& \mathcal{O}\left(h(\theta) - h(\hat{\theta}) ~ \big| ~ h(\theta) - h(\hat{\theta})\right) \nonumber \\
&=& \sum_i \lambda_i (\Delta \mu_i)^2 < \chi_k^2 
\end{eqnarray}
for two waveforms parametrized by the vector parameter $\theta = \{ \theta_i \}$ defined earlier. 

From the relative size of their eigenvalues $\lambda_i$, we can conclude which principal components affect the waveform most strongly and yield the most accurately measurable parameter combinations in both vacuum and environmentally-dephased cases. The bigger the eigenvalue, the more tightly constrained the parameter combination is, quantified by the confidence intervals $\mu_i$~\citep{Ohme2013}.

\section{Measurable Parameter Combinations}
\label{sec:measurable}

% Section 3 before 3.1 - Add “overview” - a few sentences describing what the results will be & what they mean (eigenvalues and confidence interval).

In this section, we present results of the Fisher matrix PCA applied to the vacuum EMRI waveform (Sec.~\ref{subsec:results_vac}) and the environmentally-dephased EMRI waveform (Sec.~\ref{subsec:results_env}). For each case, we report the eigenvalues $\lambda_i$ and confidence intervals $\Delta\mu_i$ of the identified principal components $\mu_i$. Recall that the relative size of the eigenvalue indicates the strength of the principal component; the bigger it is, the more tightly-constrained the parameter combination is. For the physical interpretation of the principal components, we also report the contributions of the PN parameters to each.

The target system we consider is an EMRI binary with component masses $m_1 = 10^5 M_{\odot}$ and $m_2 = 10 M_{\odot}$, dimensionless spin parameters $\chi_1 = 0.8$, and $\chi_2 = 0.5$, and luminosity distance $d_L = 1~{\rm Gpc}$. For the environmentally-dephased case, we assume a constant environmental density $\rho_0 = 100~ {\rm kg~m}^{-3}$, which is in the range expected in thin accretion disks. We note that we specifically chose this system as it has been shown that the accuracy to which LISA can probe this its environmental density can reach up to dark matter levels, from a waveform signal with SNR $\sim60$~\citep{Cardoso2020}. 

%Our goal is to show that estimating this system's environmental density will be hindered by its degeneracies with the mass and spin parameters, and that estimating the PC will be more accurate.}

For these calculations, we introduce an arbitrary normalization frequency $f_0$ and use the dimensionless frequency $f/f_0$ in lieu of the GW frequency $f$ in Eqs.~\eqref{eq:htilde_vac} \& \eqref{eq:htilde_env}. We note that the results depend on the choice of $f_0$, which we have chosen here to be 0.01~Hz, to give an indication of which PN parameters are important in the LISA band.

As an example, we show how the PCA method works for the vacuum EMRI waveform and recover some meaningful results, then we show how the method works for an environmentally-dephased waveform.

\subsection{Vacuum EMRI Waveform}
\label{subsec:results_vac}

Recall that our model vacuum EMRI waveform has six PN parameters (excluding the phase and time at coalescence terms, $\Psi_5$ and $\Psi_8$, respectively) and that the principal components are linear combinations of these PN parameters. Table~\ref{table:eigvalvac} shows the eigenvalues $\lambda_i$ and the theoretical spread in the $90\%$ confidence interval $\Delta \mu_i$ for the principal components of the Fisher matrix for the vacuum EMRI waveform, $\mu_i$. Also shown are the fraction of the variance captured by each principal component. We find that the first principal component is dominant, with an eigenvalue is 3 orders of magnitude larger than that of the second principal component, and already captures almost all the original variance in the waveforms.

%The contributions of the PN parameters to the principal components are shown in Table~\ref{table:eigvecvac}. 

%In general, the number of PC required to reconstruct a certain percentage of the variance relies on the eigenvalues.

\begin{table}[htbp]
    \centering
    \begin{tabular}{|l|c|c|c|}
    \hline
    & $\lambda_i$ & $\Delta \mu_i$ & Frac. of Variance \\ 
    \hline
    $\mu_1$ & $7.43 \times 10^3$ & $0.02900$ & $9.96 \times 10^{-1}$ \\
    $\mu_2$ & $2.96 \times 10^1$ & $0.45920$ & $3.97 \times 10^{-3}$ \\
    $\mu_3$ & $5.28 \times 10^{-1}$ & $3.43919$ & $7.08 \times 10^{-5}$ \\
    $\mu_4$ & $3.03 \times 10^{-3}$ & $45.3489$ & $4.07 \times 10^{-7}$ \\
    $\mu_5$ & $1.00 \times 10^{-5}$ & $717.219$ & $1.63 \times 10^{-9}$\\
    $\mu_6$ & $2.22 \times 10^{-8}$ & $16763.5$ & $2.98 \times 10^{-12}$\\
    \hline
    \end{tabular}
    \caption{Eigenvalues $\lambda_i$, $90\%$ confidence interval $\Delta \mu_i$, and fraction of total variance explained by each principal component $\mu_i$ for the vacuum EMRI waveform.}
    \label{table:eigvalvac}
\end{table}

\begin{table}[htbp]
    \centering
    \begin{tabular}{|l|r|r|r|r|r|r|} %{c|c|c|c|c|c|c}
    \hline
    & \multicolumn{1}{|c|}{$\Psi_0$} & \multicolumn{1}{c|}{$\Psi_2$} & \multicolumn{1}{c|}{$\Psi_3$} & 
    \multicolumn{1}{c|}{$\Psi_4$} &
    \multicolumn{1}{c|}{$\Psi_6$} &
    \multicolumn{1}{c|}{$\Psi_7$} \\
    \hline
    $\mu_1$ & $-0.92$ & $-0.34$ & $-0.18$ & $-0.07$ & $0.04$ & $0.04$ \\
    $\mu_2$ & $0.36$ & $-0.57$ & $-0.52$ & $-0.29$ & $0.26$ & $0.35$ \\
    $\mu_3$ & $0.15$ & $-0.47$ & $-0.13$ & $0.09$ & $0.40$ & $-0.75$ \\
    $\mu_4$ & $-0.06$ & $0.50$ & $-0.47$ & $-0.54$ & $0.23$ & $0.43$ \\
    $\mu_5$ & $0.02$ & $-0.25$ & $0.50$ & $0.09$ & $0.75$ & $-0.33$ \\
    $\mu_6$ & $0.00$ & $-0.13$ & $0.46$ & $-0.78$ & $-0.39$ & $0.11$ \\
    \hline
    \end{tabular}
    \caption{Contributions of the PN parameters to the principal components $\mu_i$ for the vacuum EMRI waveform.}
    \label{table:eigvecvac}
\end{table}

The principal components can be written as linear combinations of the PN parameters as
\begin{equation}
    \mu_i = \sum_j \Lambda_{ij} \theta_j
    \label{eq:PCs}
\end{equation}
where $\Lambda_{ij}$ are the individual elements reported in Table \ref{table:eigvecvac}. We find that, to a good approximation, the first principal component (with the highest eigenvalue) is
\begin{equation}
    \mu_1 \approx -0.92 \Psi_0 - 0.34 \Psi_2.
    \label{eq:PC1vac}
\end{equation}
The dominant contribution is from the zero-index PN component, which is simply proportional to the chirp mass $\Psi_0 \propto \mathcal{M}^{-5/3}$. This reproduces the result of \cite{Ohme2013} that the chirp mass is the best-measured quantity. 

%It is also noted that even for higher mass ratios or other spin configurations, the dominant contribution still stems from $\Psi_0$.

\subsection{Environmentally-dephased EMRI Waveform}
\label{subsec:results_env}

We proceed to investigate the case for our environmentally-dephased target EMRI waveform. Tables~\ref{table:eigvalenv} and \ref{table:eigvecenv} show the results of the Fisher Matrix PCA; recall that the first 3 PN parameters with negative indices correspond to the effects from gravitational drag, accretion, and gravitational pull, respectively. We find that, just as in the vacuum case, the first principal component is dominant; its eigenvalue is 3 orders of magnitude larger than that of the second principal component and it captures almost all the original variance in the waveforms. In this case, however, the dominant contributions to the first principal component come from the environmental effects, with increasing magnitude for lower PN order. This result is not surprising as the negative PN terms are expected to affect the inspiral waveform at larger distances and the more negative the PN order, the larger the effect \citep{Barausse2015}. This pattern is also consistent with the results of \cite{Cardoso2020}, which treated each effect independently and found that constraints on the environmental density is tightest for gravitational drag, followed by accretion, and gravitational pull.

From these results, we propose a reparameterization of the PN parameters encapsulating the environmental effects, based on the calculated contributions to the first principal component (c.f. Table~\ref{table:eigvecenv})
\begin{equation}
\theta_{\rm env} = 0.89\Psi_{-11} + 0.43\Psi_{-9} + 0.14\Psi_{-6} \approx \mu_1.
    \label{eq:reparam}
\end{equation}
This new parameter $\theta_{\rm env}$ can be estimated with a corresponding error of $4.95 \times 10^{-4}$ at $90\%$ confidence, which is a tighter constraint than when sampling the original individual PN or physical parameters.

%$99\%$ of the variance. Meanwhile, the size of the eigenvalues emphasize the strength of each PC, in the sense that the bigger the eigenvalue, the more accurate and constrained the parameter combinations are~\citep{Ohme2013}.

%% Note comparison with vacuum case where chirp mass is first principal component?

\begin{table}[htbp]
    \centering
    \begin{tabular}{|l|c|c|c|}
    \hline
    & $\lambda_i$ & $\Delta \mu_i$ & Frac. of Variance \\
    \hline
    $\mu_1$ & $3.17 \times 10^7$ & $4.95 \times 10^{-4}$ & $9.98 \times 10^{-1}$ \\
    $\mu_2$ & $3.30 \times 10^4$ & $1.54 \times 10^{-2}$ & $1.04 \times 10^{-3}$ \\
    $\mu_3$ & $2.46 \times 10^2$ & $1.77 \times 10^{-1}$ & $7.93 \times 10^{-6}$\\
    $\mu_4$ & $8.37 \times 10^0$ & $9.64 \times 10^{-1}$ & $2.79 \times 10^{-7}$\\
    $\mu_5$ & $2.82 \times 10^{-1}$ & $5.25 \times 10^0$ & $1.04 \times 10^{-8}$\\
    $\mu_6$ & $4.03 \times 10^{-3}$ & $4.40 \times 10^1$ & $1.58 \times 10^{-10}$ \\
    $\mu_7$ & $1.07 \times 10^{-5}$ & $8.52 \times 10^2$ & $4.64 \times 10^{-13}$\\
    $\mu_8$ & $3.37 \times 10^{-8}$ & $1.52 \times 10^4$ & $1.60 \times 10^{-15}$\\
    $\mu_9$ & $2.25 \times 10^{-11}$ & $5.89 \times 10^5$ & $1.13 \times 10^{-18}$\\
    \hline
    \end{tabular}
    \caption{Eigenvalues $\lambda_i$, $90\%$ confidence interval $\Delta \mu_i$, and fraction of total variance explained by each principal component $\mu_i$ for the EMRI waveform environmentally-dephased for gravitational drag, accretion, and gravitational pull effects.}
    \label{table:eigvalenv}
\end{table}

\begin{table*}[tp]
    \centering
    \begin{tabular}{|l|r|r|r|r|r|r|r|r|r|} %{c|c|c|c|c|c|c}
    \hline
    & \multicolumn{1}{|c|}{$\Psi_{-11}$} &
    \multicolumn{1}{|c|}{$\Psi_{-9}$} &
    \multicolumn{1}{|c|}{$\Psi_{-6}$} &
    \multicolumn{1}{|c|}{$\Psi_0$} & \multicolumn{1}{c|}{$\Psi_2$} & \multicolumn{1}{c|}{$\Psi_3$} & 
    \multicolumn{1}{c|}{$\Psi_4$} &
    \multicolumn{1}{c|}{$\Psi_6$} &
    \multicolumn{1}{c|}{$\Psi_7$} \\
    \hline
    $\mu_1$ & $0.89$ & $0.43$ & $0.14$ & $0.01$ & $0.00$ & $0.00$ & $0.00$ & $-0.00$ & $-0.00$ \\
    $\mu_2$ & $0.40$ & $-0.60$ & $-0.66$ & $-0.17$ & $-0.08$ & $-0.04$ & $-0.02$ & $0.01$ & $0.01$ \\
    $\mu_3$ & $0.17$ & $-0.48$ & $0.32$ & $0.62$ & $0.39$ & $0.26$ & $0.12$ & $-0.09$ & $-0.12$ \\
    $\mu_4$ & $0.12$ & $-0.43$ & $0.59$ & $-0.18$ & $-0.36$ & $-0.33$ & $-0.20$ & $0.22$ & $0.32$ \\
    \hline
    \end{tabular}
    \caption{Contributions of the PN parameters to the first four principal components $\mu_i$ for the EMRI waveform environmentally-dephased for gravitational drag, accretion, and gravitational pull effects (which correspond to the first 3 negative-indexed PN parameters, respectively).}
    \label{table:eigvecenv}
\end{table*}

Next, we interpret the first principal component $\mu_1$ as a function of physical parameters so we can investigate its contours in relevant physical parameter spaces. To do this, we first express it as a linear combination of PN parameters according to Eq.~\eqref{eq:PCs}, then replace each PN parameter with the phase expansion term, which depends on $M$, $\eta$, $\chi$, and $\rho_0$ according to Eq.~\eqref{eq:IMRPhenomB_phase}. 
 
\begin{figure}[h!]
    \centering
    \includegraphics[width=0.8\linewidth]{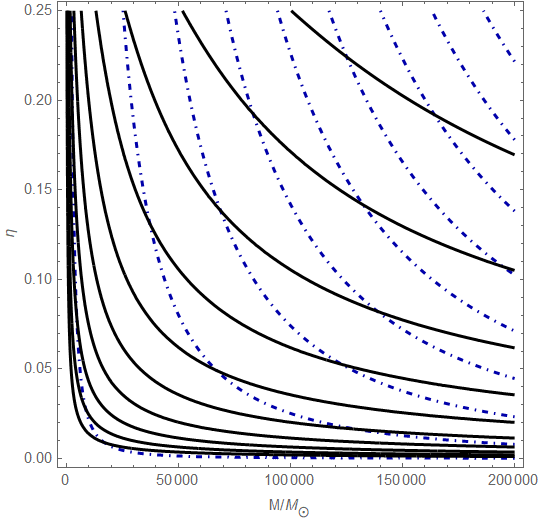}
    \caption{Contours of the (transformed) first principal component $\mu_1' = 1/\mu_1^2$ corresponding to values from $1.41 \times 10^{-20}$ to $1.41 \times 10^{-2}$ in equally-spaced steps in log scale of $2$ dex, increasing from top right to bottom left. Blue dot-dashed curves are contours of constant chirp mass $\mathcal{M} = \{1.00 \times 10^3, 1.10 \times 10^4, ..., 1.01 \times 10^5\}$ $M_\odot$, in steps of $10^4 M_\odot$.}
    \label{fig:Metaaxis}
\end{figure}

Figure~\ref{fig:Metaaxis} shows contours of the transformed variable $\mu_1' = 1/\mu_1^2$ in the $M$-$\eta$ parameter space in log steps of $10^2$ (with $\chi$ and $\rho_0$ set to the fiducial values). We see how this measurable parameter combination differs from the chirp mass, whose contours are shown in the figure as blue dot-dashed curves. Note that the $\mu_1'$ contours have less space in between them when it comes to the less massive and more asymmetric part of the parameter space, which means it can track these kinds of binaries with more precision compared to the use of chirp mass to describe a system. 

%Given the dominance of the first principal component, systems with $M$ and $\eta$ combinations along a given contour essentially have indistinguishable signatures in the LISA band.

%\textcolor{red}{This figure shows how statistical uncertainties show up for the environmentally-dephased waveform: within each contour, the same template is calculated for each pair of $M-\eta$. As such, less parameters along these contours can be used~\citep{Ohme2013,Owen1996,Baird2013}.}

% \textit{[Note to Marco: Please make the ff changes to the Figure: (1) Remove title (2) Remove top axis labels (3) Ticks in horizontal axis must be regular, first tick = 0 instead of 10; also add minor ticks and make the exactly the same as in Figure 2 (4) Change colors: make solid curves BLACK and dot-dashed curves blue (5) Change x-axis label to $M/M_\odot$.]}

\begin{figure}[h!]
    \centering
    \includegraphics[width=0.8\linewidth]{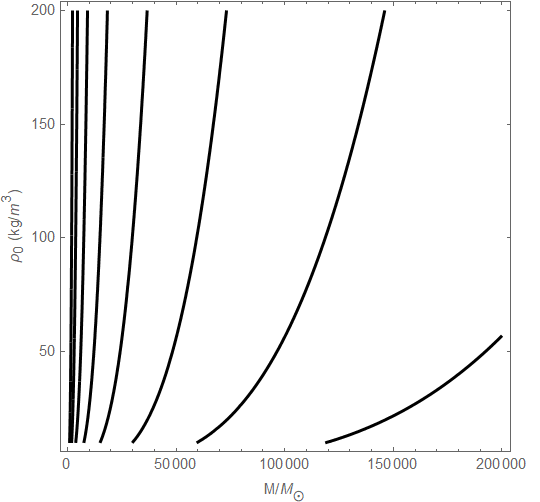}
    \caption{Contours of the transformed first principal component $\mu_1'' = 1/\mu_1$ corresponding to values from $\mu_1'' = -1.16 \times 10^{-21}$ to $-1.16 \times 10^{-14}$ in equally-spaced steps in log scale of $1$ dex, increasing from left to right.}
    \label{fig:Mrhoaxis}
\end{figure}

Finally, we investigate the dependence of $\mu_1$ on the variable of interest, environmental density. Figure~\ref{fig:Mrhoaxis} shows contours of the transformed variable $\mu_1'' = 1/\mu_1$ in the $M$-$\rho_0$ parameter space (with $\chi$ and $\eta$ set to the fiducial values). We find that for $M \lesssim 10^4 M_\odot$, $\mu_1''$ largely traces $M$ and does not constrain $\rho_0$. For larger $M$, there is a shallower slope in the contours indicating some degeneracy between $\rho_0$ and $M$. 

Note that while the reported values and contours in the physical parameter space are specific to the selected target system, the result that the negative PN orders will dominate in the inspiral range applies in the general case. Hence, the dominant measurable parameter combinations derived from PCA will be sensitive to the environmental effects and relevant for strategies in constraining the environments around EMRIs. % talk about general result of envi fx dominance due to negative PN orders / \citep{Cardoso2020}

%\textcolor{red}{This might affect some astrophysical assumptions regarding the observed system, noting that there is a correlation between central black hole mass and supported environmental density. (Since the analysis is done for an EMRI target, the total mass is dominated by the primary central black hole.)}

% Discussion on what would still hold in the general case — ie the environment effect terms are still expected to dominate (I think). We can also mention that though we leave the detailed analysis for future work, we can sketch out how it would be done (applying the current methodology for ensemble of systems).  

% \textit{[Note to Marco: Please make the ff changes to the Figure: (1) Remove title (2) Remove dashed curves (3) Change y-axis units to kg m$^{-3}$ and y-axis label to $\rho_0 {\mbox (kg/m}^{-3})$ (4) Change color to BLACK (5) Change x axis label to $M/M_\odot$ (6) Show $\mu_1$ only (remove $\mu_2$).]}

\section{Summary and Conclusions}
\label{sec:conclu}

In this work, we have expanded the application of the Fisher Matrix PCA method to investigate measurable parameter combinations and degeneracies for gravitational-wave signals from EMRIs in non-trivial astrophysical environments. Following \cite{barausse2014} and \cite{Cardoso2020}, we modeled environmental effects from gravitational drag, accretion, and gravitational pull as a dephasing of vacuum gravitational waveforms, which appear as negative-indexed PN parameters in a post-Newtonian waveform approximation. Our results show that the inspiral phase of the binary’s orbit is dominated by the environmental effects, led by gravitational drag, which has the most negative PN order. We obtain a new parameter combination (given by Eq.~\eqref{eq:reparam}) that can be used in lieu of the original individual parameters, for more efficient stochastic sampling for parameter estimation. 

%We build on this result by providing parameter combinations in the form of principal components, which has the advantage of yielding more accurate measurements rather than using the individual physical parameters. The degeneracies in the physical parameter space were also considered, since it shows which parts of the parameter space may be more useful for these parameter combinations.}

% add paragraph to highlight results w/respect to environmental effects - summarize “Table 4 result”. 

% & applications/how this can be used - emphasize advantage of PCs:

The present analysis is limited by several factors: applying the Fisher matrix PCA on a single EMRI target with a high SNR, ignoring the binary dynamics' effects on the environment, and choosing a post-Newtonian waveform approximant for simplicity. More accurate waveform approximants for EMRIs are required, especially those that account the environmental effects \citep{cardoso2022gravitational}. These waveforms may include additional information on the binary \citep{Chua2020, Chua2021}, higher order modes \citep{Bustillo2016}, or non-trivial eccentricities \citep{Favata2022}, and thus can be used to modify the parameter combinations. We also used a simplified astrophysical environment -- a proper investigation of the different environmental configurations should also take into account geometry and accretion rate as a parameter (in addition to environmental density) as in \cite{Kocsis2011} and \cite{Caputo2020}. 

% \textcolor{red}{Add here: Recommended future work is to expand the analysis to investigate ensemble of target EMRI systems... dependence on mass ratios, spins, SNR.}
As noted earlier, our analysis indicates that the dominant measurable parameter combinations will be sensitive to environmental effects and relevant for strategies in constraining the environments around EMRIs in the general case. The analysis can also be expanded to check the physical parameter dependence of the relative errors of the principal components, analogous to Figure 4 of \cite{Pai2013}. This will help us understand the behavior for other interesting systems in the LISA range such as the intermediate-mass ratio inspiral (IMRI) binaries and the massive (supermassive) black hole binaries (SMBHB).

% add a sentence on limitation of single system; further work can investigate the general case.
Finally, we note that this approach has the potential to be expanded to consider implications for tests of gravity as the PCA method can readily be applied using the parametrized post-Einsteinian framework for alternate theories of gravity \citep{yunes2009ppE}.

\section*{Acknowledgments}
The authors would like to thank Ian Vega, Andrea Maselli, and Reginald Christian Bernardo for the helpful discussions over the course of this study.

%\appendix

%\section{Sample Appendix Section}

\bibliographystyle{elsarticle-harv} 
\bibliography{cas-refs}





% \end{thebibliography}
\end{document}